\documentclass[aip,dcolumn,ajp,reprint]{revtex4-1}
\usepackage{dcolumn}
\usepackage{amsmath,amssymb}
\usepackage{graphicx,times}
\newcommand{\fracd}[2]{\cfrac{\mathrm{d} #1}{\mathrm{d} #2}} 

\def\sub#1{_{\rm #1}}

\def\ee{\mathrm e}
\def\ii{\mathrm i}
\def\cc{\mathrm{c.c.} }
\def\U#1{\,\mathrm{#1}}

\begin{document}
\title{
Efficient second harmonic generation in a metamaterial with two
resonant modes coupled through two varactor diodes
\vspace{3mm}
}
\author{Toshihiro Nakanishi}
\email{t-naka@kuee.kyoto-u.ac.jp}
\author{Yasuhiro Tamayama}
\author{Masao Kitano}
\affiliation{Department of Electronic Science and Engineering, Kyoto University, Kyoto 615-8510, Japan}
\date{\today}

\begin{abstract}
 We present an effective method to generate second harmonic (SH) waves using
 nonlinear metamaterial composed of coupled split ring resonators (CSRRs)
 with varactor  (variable capacitance) diodes.
 The CSRR structure has two resonant modes: a symmetric mode that
 resonates at the fundamental frequency and an
 anti-symmetric mode that
 resonates at the SH frequency.
 Resonant fundamental waves in the symmetric mode 
generate resonant SH waves in the anti-symmetric mode. 
The double resonance contributes to effective SH radiation.
 In the experiment,
we observe $19.6\U{dB}$ enhancement in the SH radiation
in comparison with the nonlinear metamaterial that resonates only for the 
fundamental waves. 

\end{abstract}
\pacs{
78.67.Pt,
42.65.Ky,
41.20.-q
}
\keywords{metamaterials, nonlinear optics, second harmonic generation}
\maketitle

Metamaterials, which are composed of artificial
sub-wavelength structures, exhibit
extraordinary electromagnetic properties.
Numerous studies have  focused on the  linear
response characteristics of metamaterials.
Recent studies have also reported the development of nonlinear 
media and the control of the nonlinear properties of the metamaterials 
through the introduction of nonlinear elements 
into the constituents of  metamaterial.
High nonlinearity can be achieved in
resonant-type metamaterials such as split ring resonators (SRRs) 
 by placing nonlinear elements  at  locations
where the electric (or magnetic) field is concentrated 
due to the resonance effect. \cite{pendry}
Nonlinear metamaterials have been studied for
property tuning, \cite{shadrivov,wang,powell}
frequency mixing, \cite{Huang2011} 
imaging  beyond diffraction limit, \cite{Wang2011} and
developing bistable media. \cite{wang,Yang2010, Chen2011}
Several kinds of nonlinear metamaterials
for generating second harmonic (SH) waves
have been reported. 
Most of the metamaterials are designed such that they resonate with
the incident waves or the fundamental
waves. \cite{klein,klein07,shadrivov,kim,bias, Rose2011}
These are called singly resonant metamaterials.
If the structure has resonant modes not only for 
the fundamental frequency but also for the SH frequency,
SH radiation could be enhanced \cite{gorkunov,kanazawa2011}.
In this paper, we propose a method to implement
metamaterials satisfying the doubly resonant condition,
or a doubly resonant metamaterial,
using coupled split ring resonators (CSRRs) with two varactor diodes,
which generate  SH waves due to nonlinearity.
There are two resonant modes in the CSRR structure:
one for the fundamental waves; the other for the SH waves.
They are coupled owing to the nonlinearity of the diodes.
The SH oscillation excited in the varactors
directly excites the SH resonance mode through the nonlinearity-assisted
coupling.
Second harmonic generation through the CSRRs
is more efficient than that through the previously proposed metamaterial
with two resonant modes,
where SH oscillation indirectly excites the resonant mode 
designed for the SH
waves through magnetic coupling. \cite{kanazawa2011}
We demonstrate  experiments in the microwave region to 
estimate the SHG efficiency of the CSRRs,
comparing the singly resonant metamaterial.

\begin{figure}[b]
 \includegraphics[scale=1]{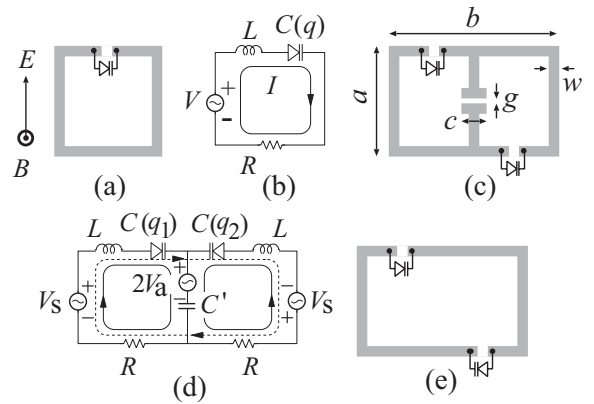}
 \caption{(a) Varactor-loaded split ring resonator and (b) its circuit model. 
 (c) Coupled split ring resonator and (d) its circuit model. (e)
 Split ring resonator as the reference metamaterial.}
 \label{fig:structure}
\end{figure}

The varactor loaded split ring resonator (SRR) shown 
in Fig.~\ref{fig:structure}(a)
is a typical example of singly resonant nonlinear metamaterials.
It can be modeled as 
a series resonant circuit composed of an inductor, varactor, and resistor,
as shown in Fig.~\ref{fig:structure}(b). 
The electromotive force induced by the external magnetic field $B$
is represented by the voltage source $V(t) = V_0  \cos \omega t$ in the circuit model.
If we consider the varactor as a linear capacitor,
the circuit is a simple harmonic oscillator driven by the external force
and the voltage across the  capacitor, $v_{C}$, reaches 
a maximum at the resonant angular frequency $\omega_0=1/\sqrt{LC}$. 
However, the varactor has nonlinearity in the capacitance
 and its voltage $v_C$ with respect to
the charge $q$ can be represented as
$v_{C}=q/C(q)=q/C+\alpha q^2$.
An anharmonic term $\alpha q^2$ contributes to the generation of second-
or higher-order harmonic waves.
 Using the perturbation method under a weak nonlinearity condition, \cite{poutrina} 
 the amplitude of the SH oscillation is obtained as
 \begin{equation}
  \tilde{q} (2 \omega) =
  \frac{\alpha V_0^2}{\omega^3
   Z(2\omega) Z(\omega)^2 } \label{eq:1}, 
 \end{equation}
where $Z(\omega)=R-\mathrm{i}\left[\omega L-1/\left(\omega C\right)\right]$ is the 
 impedance of the circuit. 
 When the circuit resonates at $\omega = \omega_0$, 
 $\left|Z(\omega)\right|$
 takes a minimum value and $\left|\tilde{q} (2 \omega)\right|$ 
 is maximized. 
 As a result, the enhanced SH signal is radiated from the singly resonant
 metamaterial. 
Here, we can expect that 
 $\left|\tilde{q} (2 \omega)\right|$ is further enhanced by also
 reducing the impedance at $2 \omega$,
i.e.~$\left|Z(2\omega)\right|$.
This implies that 
more efficient SHG can be achieved when the metamaterial
resonates for both the fundamental and SH frequencies.

For further enhancement of the SH radiation,
we introduce the CSRR shown in
Fig.~\ref{fig:structure}(c).
Two ring structures share a gap with a capacitance $C^\prime$
at the center of the structure.
Figure \ref{fig:structure}(d) represents the circuit model 
of the CSRR.
It should be noted   that
there are two diodes  oppositely directed on the outer ring.
We denote the charges of the varactors as $q_1$ and $q_2$.
In terms of the variables,
$q\sub{s} =q_1+q_2$ and $q\sub{a}=q_1-q_2$, 
the equations of motions are expressed as
\begin{eqnarray}
&& L \fracd{^2 q\sub{s}}{t^2} + R \fracd{q\sub{s}}{t} +\frac{q\sub{s}}{C}
 + \alpha q\sub{s} q\sub{a}  = 2 V\sub{s} \cos \omega t, \label{eq:sym}\\
&& L \fracd{^2 q\sub{a}}{t^2} + R \fracd{q\sub{a}}{t} + \frac{q\sub{a}}{C\sub{a}}
 + \frac{\alpha}{2} q\sub{s}^2  + \frac{\alpha}{2} q\sub{a}^2
 = 2 V\sub{a} \cos \omega t, \label{eq:asym}
\end{eqnarray}
where $1 / C\sub{a}= 1/C + 2/ C^\prime$.
The excitation voltage $V\sub{s}$ is induced by the magnetic flux
through the rings and  voltage $V\sub{a}$ is induced by the electric field 
at the central gap of the structure. 
Equations (\ref{eq:sym}) and (\ref{eq:asym}) represent that the symmetric
current (dashed line)
and anti-symmetric current (solid line) form two resonant modes with 
 resonant angular frequencies of $\omega\sub{s}=1/\sqrt{LC}$ and
$\omega\sub{a}=1/\sqrt{LC\sub{a}}$, respectively; these modes are coupled through the nonlinearity $\alpha$.
We set the gap capacitance  $C^\prime=2C/3$, so that
the doubly resonant condition $\omega\sub{a}=2\omega\sub{s}$ is satisfied.

We will solve $q\sub{s}$ and $q\sub{a}$ in the presence of a small
nonlinearity $\alpha$,
using a perturbative approach.
We assume that the solution for $\alpha=0$ is given by 
$q_i^{(0)} = \tilde{q}_i^{(0)} (\omega) \, \ee^{- \ii \omega
t} +\cc (i={\rm s, a})$.
From Eqs.~(\ref{eq:sym}) and (\ref{eq:asym}),
the amplitudes of $q\sub{s}^{(0)}$ and $q\sub{a}^{(0)}$ oscillating at $\omega$
are written as
 $\tilde{q}_i^{(0)}(\omega) = \frac{V_i}{\omega Z_i (\omega) }
 \ (i= {\mathrm s , a}),$
where $Z\sub{s}(\omega)=R-\mathrm{i}\left[\omega L-1/\left(\omega
C\right)\right]$ and $Z\sub{a}(\omega)=R-\mathrm{i}\left[\omega L-1/\left(\omega
C\sub{a}\right)\right]$ are the impedances for the symmetric  and
anti-symmetric modes, respectively.
When the excitation angular frequency $\omega$ is tuned close to $\omega\sub{s}$,
$|\tilde{q}^{(0)}\sub{s}|$ becomes large due to  resonance.
On the other hand, the anti-symmetric mode is hardly excited; $\tilde{q}^{(0)}\sub{a} \sim 0$.

The first-order solution of $q\sub{a}$ with respect to 
$\alpha$ satisfies
\begin{equation}
 L \fracd{^2 q^{(1)}\sub{a}}{t^2} + R \fracd{q^{(1)}\sub{a}}{t} 
+ \frac{q^{(1)}\sub{a}}{C\sub{a}}
 + \frac{\alpha}{2} \{q^{(0)}\sub{s}\}^2  
 = 0. \label{eq:asym2}
\end{equation}
It is found that
the resonant oscillation $q^{(0)}\sub{s}$ in the symmetric mode
induces the SH current in the
anti-symmetric mode through the last term $\frac{\alpha}{2}
\{q^{(0)}\sub{s}\}^2 $.
The nonlinearity of the varactor results in a
 coupling between the two resonant modes.
It should be noted that if the diodes are arranged in the same direction 
on the outer ring, the fundamental current and the 
induced SH current  flow in the
symmetric mode, which resonates only for the fundamental wave.
From Eq.~(\ref{eq:asym2}), the amplitude of the SH oscillation  is obtained as
\begin{eqnarray}
 \tilde{q} \sub{a} (2 \omega) &=&\frac{\alpha V\sub{s}^2}{2 \omega^3
  Z\sub{a} (2 \omega)
 Z\sub{s} (\omega)^2
  }. \label{eq:shg}
\end{eqnarray}
It is easily deduced from Eq.~(\ref{eq:sym}) that there 
are no SH waves in the
symmetric mode. 
This is because the induced SH electromotive voltages in the two diodes
cancel each other due to the anti-symmetric arrangements
of the diodes for the symmetric mode.
The SH oscillation excited only in the anti-symmetric mode forms an 
electric dipole with 
magnitude  $\tilde{p}= \tilde{q}\sub{a} (2 \omega) \bar{g}$, where
$\bar{g}$ is an effective dipole length and is determined by 
the current or charge distributions. The electric dipole oscillation 
contributes to efficient SH radiation.
Both  $|Z\sub{s} (\omega)|$ and $|Z\sub{a} (2
\omega)|$ in Eq.~(\ref{eq:shg}) are  small,
because of the doubly resonant condition, $\omega=\omega\sub{s}=\omega\sub{a}/2$.
Consequently, the amplitude of the second harmonic oscillation 
can be significantly enhanced in comparison to the singly resonant
metamaterial, where the SH current has to flow through the high
impedance $|Z(2 \omega)|$ in Eq.~(\ref{eq:1}).

We fabricated the CSRR
illustrated in Fig.~\ref{fig:structure}(c) with
$35\U{\mu m}$-thick copper film on 
a polyphenylene ether (PPE) substrate with a thickness of $0.8\U{mm}$.
The dimensions are $a=14\U{mm}$, $b=24\U{mm}$, $c=4\U{mm}$, $g=0.5\U{mm}$,
and $w=1\U{mm}$.
For comparison, we also prepared a reference metamaterial, shown
in Fig.~\ref{fig:structure}(e),
which has the same structure, except for the absence of the central structure
and the direction of the diodes.
The loop current in this structure also resonates at the same fundamental frequency
as that of the CSRR, and the SH electromotive voltages generated in the two
varactors contribute to the loop current, which does not resonate.
Therefore, the ratio of SHG efficiency between CSRRs and SRRs 
can be interpreted as the enhancement factor
owing to the resonance effect for the SH waves.

\begin{figure}[]
 \includegraphics[scale=1]{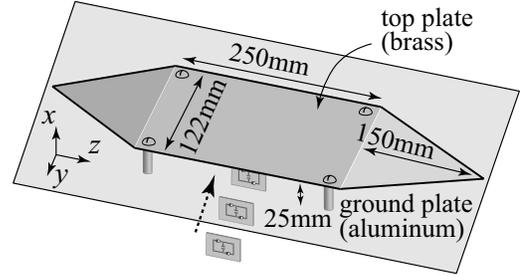}
 \caption{Schematic of stripline TEM-cell waveguide.}
 \label{fig:trans}
\end{figure}

Before SHG demonstration,
we conducted transmission measurements to identify the resonant modes
and resonant frequency of the metamaterials.
We used a waveguide called a stripline TEM-cell, as shown in 
Fig.~\ref{fig:trans}. \cite{Collin, poutrina}
The waveguide supports a TEM mode propagating in the $z$ direction.
The tapered structures ensure that the wave impedance
is maintained at $50\U{\Omega}$ along the waveguide
and the tips of the top plate are connected to the input and output
ports,
which have SMA connectors.
We placed three SRRs or CSRRs at intervals of $5\U{cm}$ 
inside the waveguide,
as shown in Fig.~\ref{fig:trans}.
In the transmission spectra,
we observed two resonant dips at $1.36\U{GHz}$ and 
$2.72\U{GHz}$ for the CSRRs
and  a  dip at $1.36\U{GHz}$ for the SRRs.
Thus, it is found that the common dips at $1.36\U{GHz}$
correspond to  resonance in the symmetric (or loop-current) mode
and the higher resonant mode at $2.72\U{GHz}$ 
for the CSRRs is the anti-symmetric mode.
This CSRR structure satisfies
the doubly resonant condition, $2 \omega\sub{s}=\omega\sub{a}$.

\begin{figure}[]
 \includegraphics[scale=1]{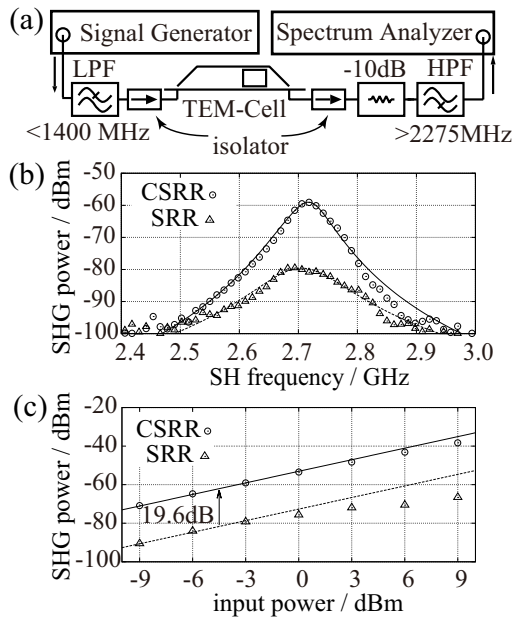}
 \caption{(a) Experimental setup for SHG measurements. (b) SHG power for
 $-3\U{dBm}$ input. (c) SHG peak power as a function of input power.}
 \label{fig:shg}
\end{figure}

Figure \ref{fig:shg}(a) shows the experimental setup 
to measure the SH power generated in the metamaterials.
A low pass filter (LPF) was used to suppress residual harmonics
from the signal generator (Agilent N5183A).
The output signal containing the SH wave passes through a high pass filter
(HPF) to reject the fundamental wave and is sent to the spectrum analyzer
(ADVANTEST U3751).
While sweeping the frequency of the input signal,
the power density at the SH frequency, or SHG power, was observed.
Figure \ref{fig:shg}(b) shows the SHG power obtained for the CSRR
(circles) and SRR (triangles) metamaterials for
an input power of $-3\U{dBm}$.
The peaks are found around $2.7\U{GHz}$ for both cases,
and the peak value for the CSRRs is  higher than that of the SRRs
by about $20\U{dB}$.
This is  clear evidence that SHG in the CSRR metamaterial 
is significantly enhanced by the resonance at the SH frequency.
As shown in Fig.~\ref{fig:shg}(b),
we fit experimental data with theoretical curves
(a solid line for the CSRR and dashed line for the SRR),
which can be derived from the fact that
the SHG power is quadratically proportional to
Eq.~(\ref{eq:1}) for the SRR and Eq.~(\ref{eq:shg}) for the CSRR.
The fitting curves reproduce the experimental results well.
Figure~\ref{fig:shg}(c) shows the peak values of the SHG spectra.
The circles and triangles correspond to the data for the CSRR and
SRR, respectively.
For weak input power,
both SHG spectra exhibit quadratic dependence, which is represented 
by the solid line for the CSRRs and the dashed line for the SRRs.
In this region,
the second order perturbation is valid,
whereas for greater input power,
third-order nonlinearity should be taken into consideration.
In fact, in the higher input power region, where the SHG spectra 
are below the solid or dashed lines,
the resonance linewidth broadens 
and a small resonant frequency shift is observed
due to the self-phase modulation.
From the above reasons, it is appropriate to estimate the SHG enhancement
of the CSRR  for weak input power without third- or higher-order nonlinearities.
The difference between the solid line and  dashed line
is $19.6\U{dB}$, which corresponds to the enhancement factor 
owing to the resonance effect for the SH wave.

As for other types of doubly resonant metamaterials,
magnetically coupled split ring resonators have been theoretically
analyzed. \cite{gorkunov}
Magnetically coupled cut-wire resonators has been studied 
and $6.6\U{dB}$ enhancement of SHG has been achieved. \cite{kanazawa2011}
Both of the metamaterials are
composite metamaterial composed of two resonant structures,
a primary resonator and secondary resonator,
which are magnetically coupled.
The SH waves generated in the primary resonator 
indirectly excite the secondary resonator  through  magnetic coupling.
Roughly speaking,
the efficiency of the SHG for these metamaterials 
in the strong coupling limit could approach that of the CSRR metamaterial,
where the SH oscillation directly excites the resonant mode 
owing to the arrangement of the nonlinear elements.
However, 
it is actually difficult to attain strong magnetic coupling 
 with finite element dimensions
and also difficult to optimize the parameters of the metamaterials,
considering the resonant frequency shifts induced by strong coupling.
The CSRR metamaterial attains
a much higher enhancement factor,
just by designing the structures to satisfy $\omega\sub{a}=2\omega\sub{s}$.

In this paper, we proposed a method to enhance SHG with CSRR
metamaterial.
The arrangement of the nonlinear elements 
induces coupling between
the two resonant modes, which satisfies the doubly resonant condition.
Owing to the direct excitation in the resonant mode 
for SH waves, we attained significant enhancement of 
up to two orders of magnitude.
The precise and quantitative evaluation of SHG efficiencies and optimization of the CSRR structure are left as future problems.
We expect  that
 metamaterials with  nonlinearity-assisted coupling
could be employed for investigating  other types of nonlinear phenomena.

The present research was supported  by Grants-in-Aid for
 Scientific Research Nos. 22109004 and 22560041,
 by the Global COE program of Kyoto University, 
and by a research grant from The Murata Science Foundation. 
One of the authors (Y.T.) would like to acknowledge the support of a 
Research Fellowship of the Japan Society for the Promotion of Science for Young Scientists.

%
%
%
%
%


\begin{thebibliography}{17}%
\makeatletter
\providecommand \@ifxundefined [1]{%
 \@ifx{#1\undefined}
}%
\providecommand \@ifnum [1]{%
 \ifnum #1\expandafter \@firstoftwo
 \else \expandafter \@secondoftwo
 \fi
}%
\providecommand \@ifx [1]{%
 \ifx #1\expandafter \@firstoftwo
 \else \expandafter \@secondoftwo
 \fi
}%
\providecommand \natexlab [1]{#1}%
\providecommand \enquote  [1]{``#1''}%
\providecommand \bibnamefont  [1]{#1}%
\providecommand \bibfnamefont [1]{#1}%
\providecommand \citenamefont [1]{#1}%
\providecommand \href@noop [0]{\@secondoftwo}%
\providecommand \href [0]{\begingroup \@sanitize@url \@href}%
\providecommand \@href[1]{\@@startlink{#1}\@@href}%
\providecommand \@@href[1]{\endgroup#1\@@endlink}%
\providecommand \@sanitize@url [0]{\catcode `\\12\catcode `\$12\catcode
  `\&12\catcode `\#12\catcode `\^12\catcode `\_12\catcode `\%12\relax}%
\providecommand \@@startlink[1]{}%
\providecommand \@@endlink[0]{}%
\providecommand \url  [0]{\begingroup\@sanitize@url \@url }%
\providecommand \@url [1]{\endgroup\@href {#1}{\urlprefix }}%
\providecommand \urlprefix  [0]{URL }%
\providecommand \Eprint [0]{\href }%
\providecommand \doibase [0]{http://dx.doi.org/}%
\providecommand \selectlanguage [0]{\@gobble}%
\providecommand \bibinfo  [0]{\@secondoftwo}%
\providecommand \bibfield  [0]{\@secondoftwo}%
\providecommand \translation [1]{[#1]}%
\providecommand \BibitemOpen [0]{}%
\providecommand \bibitemStop [0]{}%
\providecommand \bibitemNoStop [0]{.\EOS\space}%
\providecommand \EOS [0]{\spacefactor3000\relax}%
\providecommand \BibitemShut  [1]{\csname bibitem#1\endcsname}%
\let\auto@bib@innerbib\@empty
\bibitem [{\citenamefont {Pendry}\ \emph {et~al.}(1999)\citenamefont {Pendry},
  \citenamefont {Holden}, \citenamefont {Robbins},\ and\ \citenamefont
  {Stewart}}]{pendry}%
  \BibitemOpen
  \bibfield  {author} {\bibinfo {author} {\bibfnamefont {J.~B.}\ \bibnamefont
  {Pendry}}, \bibinfo {author} {\bibfnamefont {A.~J.}\ \bibnamefont {Holden}},
  \bibinfo {author} {\bibfnamefont {D.~J.}\ \bibnamefont {Robbins}}, \ and\
  \bibinfo {author} {\bibfnamefont {W.~J.}\ \bibnamefont {Stewart}},\ }\href
  {\doibase 10.1109/22.798002} {\bibfield  {journal} {\bibinfo  {journal} {IEEE
  Trans. Microwave Theory Tech.}\ }\textbf {\bibinfo {volume} {47}},\ \bibinfo
  {pages} {2075} (\bibinfo {year} {1999})}\BibitemShut {NoStop}%
\bibitem [{\citenamefont {Shadrivov}\ \emph {et~al.}(2008)\citenamefont
  {Shadrivov}, \citenamefont {Kozyrev}, \citenamefont {van~der Weide},\ and\
  \citenamefont {Kivshar}}]{shadrivov}%
  \BibitemOpen
  \bibfield  {author} {\bibinfo {author} {\bibfnamefont {I.~V.}\ \bibnamefont
  {Shadrivov}}, \bibinfo {author} {\bibfnamefont {A.~B.}\ \bibnamefont
  {Kozyrev}}, \bibinfo {author} {\bibfnamefont {D.~W.}\ \bibnamefont {van~der
  Weide}}, \ and\ \bibinfo {author} {\bibfnamefont {Y.~S.}\ \bibnamefont
  {Kivshar}},\ }\href@noop {} {\bibfield  {journal} {\bibinfo  {journal} {Appl.
  Phys. Lett.}\ }\textbf {\bibinfo {volume} {93}},\ \bibinfo {pages} {161903}
  (\bibinfo {year} {2008})}\BibitemShut {NoStop}%
\bibitem [{\citenamefont {Wang}\ \emph {et~al.}(2008)\citenamefont {Wang},
  \citenamefont {Zhou}, \citenamefont {Koschny},\ and\ \citenamefont
  {Soukoulis}}]{wang}%
  \BibitemOpen
  \bibfield  {author} {\bibinfo {author} {\bibfnamefont {B.}~\bibnamefont
  {Wang}}, \bibinfo {author} {\bibfnamefont {J.}~\bibnamefont {Zhou}}, \bibinfo
  {author} {\bibfnamefont {T.}~\bibnamefont {Koschny}}, \ and\ \bibinfo
  {author} {\bibfnamefont {C.~M.}\ \bibnamefont {Soukoulis}},\ }\href@noop {}
  {\bibfield  {journal} {\bibinfo  {journal} {Opt. Express}\ }\textbf {\bibinfo
  {volume} {16}},\ \bibinfo {pages} {16058} (\bibinfo {year}
  {2008})}\BibitemShut {NoStop}%
\bibitem [{\citenamefont {Powell}, \citenamefont {Shadrivov},\ and\
  \citenamefont {Kivshar}(2009)}]{powell}%
  \BibitemOpen
  \bibfield  {author} {\bibinfo {author} {\bibfnamefont {D.~A.}\ \bibnamefont
  {Powell}}, \bibinfo {author} {\bibfnamefont {I.~V.}\ \bibnamefont
  {Shadrivov}}, \ and\ \bibinfo {author} {\bibfnamefont {Y.~S.}\ \bibnamefont
  {Kivshar}},\ }\href {\doibase 10.1063/1.3212726} {\bibfield  {journal}
  {\bibinfo  {journal} {Appl. Phys. Lett.}\ }\textbf {\bibinfo {volume} {95}},\
  \bibinfo {pages} {084102} (\bibinfo {year} {2009})}\BibitemShut {NoStop}%
\bibitem [{\citenamefont {Huang}\ \emph {et~al.}(2011)\citenamefont {Huang},
  \citenamefont {Rose}, \citenamefont {Poutrina}, \citenamefont {Larouche},\
  and\ \citenamefont {Smith}}]{Huang2011}%
  \BibitemOpen
  \bibfield  {author} {\bibinfo {author} {\bibfnamefont {D.}~\bibnamefont
  {Huang}}, \bibinfo {author} {\bibfnamefont {A.}~\bibnamefont {Rose}},
  \bibinfo {author} {\bibfnamefont {E.}~\bibnamefont {Poutrina}}, \bibinfo
  {author} {\bibfnamefont {S.}~\bibnamefont {Larouche}}, \ and\ \bibinfo
  {author} {\bibfnamefont {D.~R.}\ \bibnamefont {Smith}},\ }\href {\doibase
  10.1063/1.3592574} {\bibfield  {journal} {\bibinfo  {journal} {Appl. Phys.
  Lett.}\ }\textbf {\bibinfo {volume} {98}},\ \bibinfo {pages} {204102}
  (\bibinfo {year} {2011})}\BibitemShut {NoStop}%
\bibitem [{\citenamefont {Wang}\ \emph {et~al.}(2011)\citenamefont {Wang},
  \citenamefont {Luo}, \citenamefont {Jiang}, \citenamefont {Wang},
  \citenamefont {Huangfu},\ and\ \citenamefont {Ran}}]{Wang2011}%
  \BibitemOpen
  \bibfield  {author} {\bibinfo {author} {\bibfnamefont {Z.}~\bibnamefont
  {Wang}}, \bibinfo {author} {\bibfnamefont {Y.}~\bibnamefont {Luo}}, \bibinfo
  {author} {\bibfnamefont {T.}~\bibnamefont {Jiang}}, \bibinfo {author}
  {\bibfnamefont {Z.}~\bibnamefont {Wang}}, \bibinfo {author} {\bibfnamefont
  {J.}~\bibnamefont {Huangfu}}, \ and\ \bibinfo {author} {\bibfnamefont
  {L.}~\bibnamefont {Ran}},\ }\href {\doibase 10.1103/PhysRevLett.106.047402}
  {\bibfield  {journal} {\bibinfo  {journal} {Phys. Rev. Lett.}\ }\textbf
  {\bibinfo {volume} {106}},\ \bibinfo {pages} {047402} (\bibinfo {year}
  {2011})}\BibitemShut {NoStop}%
\bibitem [{\citenamefont {Yang}\ and\ \citenamefont
  {Shadrivov}(2010)}]{Yang2010}%
  \BibitemOpen
  \bibfield  {author} {\bibinfo {author} {\bibfnamefont {R.}~\bibnamefont
  {Yang}}\ and\ \bibinfo {author} {\bibfnamefont {I.~V.}\ \bibnamefont
  {Shadrivov}},\ }\href {\doibase 10.1063/1.3525172} {\bibfield  {journal}
  {\bibinfo  {journal} {Appl. Phys. Lett.}\ }\textbf {\bibinfo {volume} {97}},\
  \bibinfo {pages} {231114} (\bibinfo {year} {2010})}\BibitemShut {NoStop}%
\bibitem [{\citenamefont {Chen}, \citenamefont {Farhat},\ and\ \citenamefont
  {Al\`{u}}(2011)}]{Chen2011}%
  \BibitemOpen
  \bibfield  {author} {\bibinfo {author} {\bibfnamefont {P.-Y.}\ \bibnamefont
  {Chen}}, \bibinfo {author} {\bibfnamefont {M.}~\bibnamefont {Farhat}}, \ and\
  \bibinfo {author} {\bibfnamefont {A.}~\bibnamefont {Al\`{u}}},\ }\href
  {\doibase 10.1103/PhysRevLett.106.105503} {\bibfield  {journal} {\bibinfo
  {journal} {Phys. Rev. Lett.}\ }\textbf {\bibinfo {volume} {106}},\ \bibinfo
  {pages} {105503} (\bibinfo {year} {2011})}\BibitemShut {NoStop}%
\bibitem [{\citenamefont {Klein}\ \emph {et~al.}(2006)\citenamefont {Klein},
  \citenamefont {Enkrich}, \citenamefont {Wegener},\ and\ \citenamefont
  {Linden}}]{klein}%
  \BibitemOpen
  \bibfield  {author} {\bibinfo {author} {\bibfnamefont {M.~W.}\ \bibnamefont
  {Klein}}, \bibinfo {author} {\bibfnamefont {C.}~\bibnamefont {Enkrich}},
  \bibinfo {author} {\bibfnamefont {M.}~\bibnamefont {Wegener}}, \ and\
  \bibinfo {author} {\bibfnamefont {S.}~\bibnamefont {Linden}},\ }\href@noop {}
  {\bibfield  {journal} {\bibinfo  {journal} {Science}\ }\textbf {\bibinfo
  {volume} {313}},\ \bibinfo {pages} {502} (\bibinfo {year}
  {2006})}\BibitemShut {NoStop}%
\bibitem [{\citenamefont {Klein}\ \emph {et~al.}(2007)\citenamefont {Klein},
  \citenamefont {Wegener}, \citenamefont {Feth},\ and\ \citenamefont
  {Linden}}]{klein07}%
  \BibitemOpen
  \bibfield  {author} {\bibinfo {author} {\bibfnamefont {M.~W.}\ \bibnamefont
  {Klein}}, \bibinfo {author} {\bibfnamefont {M.}~\bibnamefont {Wegener}},
  \bibinfo {author} {\bibfnamefont {N.}~\bibnamefont {Feth}}, \ and\ \bibinfo
  {author} {\bibfnamefont {S.}~\bibnamefont {Linden}},\ }\href {\doibase
  10.1364/OE.15.005238} {\bibfield  {journal} {\bibinfo  {journal} {Opt.
  Express}\ }\textbf {\bibinfo {volume} {15}},\ \bibinfo {pages} {5238}
  (\bibinfo {year} {2007})}\BibitemShut {NoStop}%
\bibitem [{\citenamefont {Kim}\ \emph {et~al.}(2008)\citenamefont {Kim},
  \citenamefont {Wang}, \citenamefont {Wu}, \citenamefont {Yu},\ and\
  \citenamefont {Shen}}]{kim}%
  \BibitemOpen
  \bibfield  {author} {\bibinfo {author} {\bibfnamefont {E.}~\bibnamefont
  {Kim}}, \bibinfo {author} {\bibfnamefont {F.}~\bibnamefont {Wang}}, \bibinfo
  {author} {\bibfnamefont {W.}~\bibnamefont {Wu}}, \bibinfo {author}
  {\bibfnamefont {Z.}~\bibnamefont {Yu}}, \ and\ \bibinfo {author}
  {\bibfnamefont {Y.}~\bibnamefont {Shen}},\ }\href {\doibase
  10.1103/PhysRevB.78.113102} {\bibfield  {journal} {\bibinfo  {journal} {Phys.
  Rev. B}\ }\textbf {\bibinfo {volume} {78}},\ \bibinfo {pages} {113102}
  (\bibinfo {year} {2008})}\BibitemShut {NoStop}%
\bibitem [{\citenamefont {Wang}\ \emph {et~al.}(2009)\citenamefont {Wang},
  \citenamefont {Luo}, \citenamefont {Peng}, \citenamefont {Huangfu},
  \citenamefont {Jiang}, \citenamefont {Wang}, \citenamefont {Chen},\ and\
  \citenamefont {Ran}}]{bias}%
  \BibitemOpen
  \bibfield  {author} {\bibinfo {author} {\bibfnamefont {Z.}~\bibnamefont
  {Wang}}, \bibinfo {author} {\bibfnamefont {Y.}~\bibnamefont {Luo}}, \bibinfo
  {author} {\bibfnamefont {L.}~\bibnamefont {Peng}}, \bibinfo {author}
  {\bibfnamefont {J.}~\bibnamefont {Huangfu}}, \bibinfo {author} {\bibfnamefont
  {T.}~\bibnamefont {Jiang}}, \bibinfo {author} {\bibfnamefont
  {D.}~\bibnamefont {Wang}}, \bibinfo {author} {\bibfnamefont {H.}~\bibnamefont
  {Chen}}, \ and\ \bibinfo {author} {\bibfnamefont {L.}~\bibnamefont {Ran}},\
  }\href {\doibase 10.1063/1.3111437} {\bibfield  {journal} {\bibinfo
  {journal} {Appl. Phys. Lett.}\ }\textbf {\bibinfo {volume} {94}},\ \bibinfo
  {pages} {134102} (\bibinfo {year} {2009})}\BibitemShut {NoStop}%
\bibitem [{\citenamefont {Rose}, \citenamefont {Huang},\ and\ \citenamefont
  {Smith}(2011)}]{Rose2011}%
  \BibitemOpen
  \bibfield  {author} {\bibinfo {author} {\bibfnamefont {A.}~\bibnamefont
  {Rose}}, \bibinfo {author} {\bibfnamefont {D.}~\bibnamefont {Huang}}, \ and\
  \bibinfo {author} {\bibfnamefont {D.~R.}\ \bibnamefont {Smith}},\ }\href
  {http://arxiv.org/abs/1106.0022} {\bibfield  {journal} {\bibinfo  {journal}
  {Phys. Rev. Lett.}\ }\textbf {\bibinfo {volume} {107}},\ \bibinfo {pages}
  {063902} (\bibinfo {year} {2011})}\BibitemShut {NoStop}%
\bibitem [{\citenamefont {Gorkunov}, \citenamefont {Shadrivov},\ and\
  \citenamefont {Kivshar}(2006)}]{gorkunov}%
  \BibitemOpen
  \bibfield  {author} {\bibinfo {author} {\bibfnamefont {M.~V.}\ \bibnamefont
  {Gorkunov}}, \bibinfo {author} {\bibfnamefont {I.~V.}\ \bibnamefont
  {Shadrivov}}, \ and\ \bibinfo {author} {\bibfnamefont {Y.~S.}\ \bibnamefont
  {Kivshar}},\ }\href@noop {} {\bibfield  {journal} {\bibinfo  {journal} {Appl.
  Phys. Lett.}\ }\textbf {\bibinfo {volume} {88}},\ \bibinfo {pages} {071912}
  (\bibinfo {year} {2006})}\BibitemShut {NoStop}%
\bibitem [{\citenamefont {Kanazawa}\ \emph {et~al.}(2011)\citenamefont
  {Kanazawa}, \citenamefont {Tamayama}, \citenamefont {Nakanishi},\ and\
  \citenamefont {Kitano}}]{kanazawa2011}%
  \BibitemOpen
  \bibfield  {author} {\bibinfo {author} {\bibfnamefont {T.}~\bibnamefont
  {Kanazawa}}, \bibinfo {author} {\bibfnamefont {Y.}~\bibnamefont {Tamayama}},
  \bibinfo {author} {\bibfnamefont {T.}~\bibnamefont {Nakanishi}}, \ and\
  \bibinfo {author} {\bibfnamefont {M.}~\bibnamefont {Kitano}},\ }\href
  {\doibase 10.1063/1.3610471} {\bibfield  {journal} {\bibinfo  {journal}
  {Appl. Phys. Lett.}\ }\textbf {\bibinfo {volume} {99}},\ \bibinfo {pages}
  {024101} (\bibinfo {year} {2011})}\BibitemShut {NoStop}%
\bibitem [{\citenamefont {Poutrina}, \citenamefont {Huang},\ and\ \citenamefont
  {Smith}(2010)}]{poutrina}%
  \BibitemOpen
  \bibfield  {author} {\bibinfo {author} {\bibfnamefont {E.}~\bibnamefont
  {Poutrina}}, \bibinfo {author} {\bibfnamefont {D.}~\bibnamefont {Huang}}, \
  and\ \bibinfo {author} {\bibfnamefont {D.~R.}\ \bibnamefont {Smith}},\
  }\href@noop {} {\bibfield  {journal} {\bibinfo  {journal} {New J. Phys.}\
  }\textbf {\bibinfo {volume} {12}},\ \bibinfo {pages} {093010} (\bibinfo
  {year} {2010})}\BibitemShut {NoStop}%
\bibitem [{\citenamefont {Collin}(1991)}]{Collin}%
  \BibitemOpen
  \bibfield  {author} {\bibinfo {author} {\bibfnamefont {R.~E.}\ \bibnamefont
  {Collin}},\ }\href@noop {} {\emph {\bibinfo {title} {Field theory of guided
  waves}}}\ (\bibinfo  {publisher} {IEEE Press, New York},\ \bibinfo {year}
  {1991})\BibitemShut {NoStop}%
\end{thebibliography}
\end{document}